# Markov Property of Continuous Dislocation Band Propagation

A. Chatterjee [a,*], A. Sarkar [a], Sourav Bhattacharya [b], P. Mukherjee [a], N. Gayathri [a] and P. Barat [a]

[a] *Variable Energy Cyclotron Centre,1/AF Bidhan Nagar, Kolkata 700064, India*

[b] *Helsinki Institute for Information Technology, University of Helsinki, Finland*

*Tensile tests were carried out by deforming polycrystalline samples of substitutional Al-2.5%Mg alloy at room temperature for a range of strain rates. The Portevin-Le Chatelier (PLC) effect was observed throughout the strain rate regime. The deformation bands in this region are found to be of type A in nature. From the analysis of the experimental stress time series data we could infer that the dynamics of type A dislocation band propagation is a Markov process.*

In a certain range of strain rate and temperature, many solid solutions, both interstitial and substitutional, exhibit the phenomenon of serrated yielding, called Portevin-Le Chatelier effect (PLC) [1,2]. In polycrystals, deformed with constant imposed strain rate, the sudden plastic strain bursts give rise to stress serrations by repeated load drops. Theoretical models explain the phenomenon in terms of the Dynamic Strain Aging (DSA) due to the interaction between moving dislocations and diffusing solute atoms [3]. The repeated breakaway of dislocations from the solid clouds leads to a negative strain rate sensitivity (SRS) of the flow stress.

* Corresponding author: arnomitra@veccal.ernet.in

Consequently, the strain localizes into narrow deformation bands on the deformed material. According to the kinetics of the formation and displacement of the PLC bands three types of instabilities can be distinguished, labeled as type A, type B, and type C [4]. On increasing strain rate and decreasing temperature, one first finds randomly nucleated static bands of type C. Then comes type B band with smaller serrations where marginal spatial correlation gives the impression of hopping propagation. Finally, the continuously propagating type A band is observed. Type A band is associated with small stress drops.

Beyond its importance in metallurgy, PLC effect is a paradigm of a general class of complex nonlinear driven systems like earth quake, avalanches of magnetic vortices in superconductors etc. [5,6,7]. But in reality, it is very difficult to analyze all these phenomena. As PLC effect serves as a model system for them, the dynamical study of the PLC effect can give an insight of these complex systems as well. So this effect has drawn continuous attention. Since PLC effect is very complex, it is very difficult to extract all the information about the dynamics from the one dimensional stress-time series data. But it is possible to predict the nature of the underlying dynamics from these experimental data.

Aluminium (Al) alloys with nominal percentage of Magnesium (Mg) show the PLC effect at room temperature when the specimens are deformed at a particular range of strain rates. To study the property of the type A PLC serrations we have carried out tensile tests on flat Al-2.5%Mg alloy samples at a strain rate of $1.18\times10^{-3}s^{-1}$, where only type A serrations are observed[4]. Tensile tests were conducted on flat specimens Al-2.5%Mg alloy in a servo controlled INSTRON (model 4482) machine. The whole experiment is conducted in room temperature (300K) and stress-time data was recorded automatically at a periodic time interval of 0.05s.

The stress data obtained during deformation shows an increasing drift due to the strain hardening effect. The drift is corrected by the method of polynomial fitting and the analyses were carried out on the corrected data [8]. Fig. 1 shows a typical segment of the drift corrected stress-time curve. However, it is observed that if the experiment is performed with similar samples under identical experimental condition, the stress-time series data are not identical. This variation is due to the randomness in the initial microstructure of each sample and also the stochastic nature of the PLC dynamics. Hence, to capture the intrinsic features of the dynamics of the PLC effect we have carried out the same tensile tests with 25 identical samples and constructed the covariance matrix from these test data for further analysis to capture the features of the dynamics. To extract any dynamical features from an experimental time series, the data has to be stationary in statistical sense. As the first requirement of stationarity, the time series should cover a period of time which is much longer than the longest characteristic time scale that is relevant for the evolution of the system. Autocorrelation time can serve as such a time scale. Fig. 2 shows the autocorrelation function (ACF) of the one dimensional stress-time series data. It is seen that the autocorrelation time (i.e. the time at which the ACF meets zero) is very small compared to the total stretch of experimental time under consideration . So we may approximate the PLC effect to be a stationary process [9]. To establish the stationarity on a strong basis, we have calculated the running mean and running standard deviation of the time series data. For a stationary process, those quantities should not differ beyond their statistical fluctuations. The statistical fluctuations are extracted from the surrogate data which represent a stationary process. Here, the surrogates are constructed from the Fourier transform of the actual data by introducing random phases, keeping the amplitude constant. Fig. 3 and fig. 4 display the running mean and running standard deviation of actual time series data and its surrogates respectively. It is quite obvious from the figures that for the experimental stress-time data, those values vary within their statistical fluctuations. So the PLC time series data represents a stationary process. Fig. 5 shows the probability distribution of the stress data. The distribution is peaked and fits very well with a Gaussian distribution. The sum of any number of data points also follows a Gaussian distribution. The Kurtosis of the distribution is 0.1, which is close to the Gaussian limit (0) [10]. Moreover, fig. 6 shows that the quantile-quantile plot (q-q plot) of the empirical Gaussian data and the actual experimental data more or less follows the straight line of slope 1. All these observations confirm the fact that the PLC stress drops are the outcome of a Gaussian process.

At this point we recall an important theorem by J. Doob [11] which states that: a one-dimensional Gaussian random process will be Markovian only when the correlation function is of the form $e^{-\alpha\tau}$ . Doob's theorem can be generalized to n-dimensions. In that case, instead of a single correlation function one must deal with the covariance matrix which can be written as $C(\tau) = \exp(Q\tau)$, where Q is some matrix. From this it immediately follows that for an *n*-dimensional Gaussian Markov process the eigen value spectrum of the covariance matrix will be of an exponential form. This can be proved as follows:

Let, $C(\tau) = \exp(Q\tau)$ is the covariance matrix. Hence it can be expressed as

$$\exp(Q\tau) = I + Q\tau + \frac{Q^2\tau^2}{2!} + \ldots.$$

We now diagonalise it:

$$S^{-1}C(\tau)S = S^{-1}(e^{Q\tau})S$$

$$= S^{-1}S + \tau S^{-1}QS + \frac{\tau^2}{2!}(S^{-1}QS)(S^{-1}QS) + \ldots\ldots$$

If $\lambda_i$s be the eigen values of $C(\tau)$ and $\beta_i$s are those for Q, then from the previous equation we get:

$$\begin{bmatrix} \lambda_1 & & & & & \\ & \lambda_2 & & & & \\ & & . & & & \\ & & & . & & \\ & & & & . & \\ & & & & & \lambda_n \end{bmatrix} = I + \tau \begin{bmatrix} \beta_1 & & & & & \\ & \beta_2 & & & & \\ & & . & & & \\ & & & . & & \\ & & & & . & \\ & & & & & \beta_n \end{bmatrix} + \frac{\tau^2}{2!} \begin{bmatrix} \beta_1^2 & & & & & \\ & \beta_2^2 & & & & \\ & & . & & & \\ & & & . & & \\ & & & & . & \\ & & & & & \beta_n^2 \end{bmatrix} + \ldots\ldots$$

$$= \begin{bmatrix} e^{\beta_1\tau} & & & & & \\ & e^{\beta_2\tau} & & & & \\ & & . & & & \\ & & & . & & \\ & & & & . & \\ & & & & & e^{\beta_n\tau} \end{bmatrix}$$

Then it follows,

$$\lambda_1 = \exp(\beta_1\tau), \quad \lambda_2 = \exp(\beta_2\tau), \quad \lambda_3 = \exp(\beta_3\tau)$$

i.e the eigen value spectrum of the covariance matrix will have an exponential decay pattern for the Markov process. From the 25 data sets containing 1000 data points each, a 1000×25 matrix is generated from which the covariance matrix is constructed. Fig. 7 shows the eigen value spectrum of the covariance matrix. It is seen that the eigen values decay in an exponential manner. This proves the fact that the PLC effect at high strain rates (i.e. for type A band regime) has Markov property. Hence, the stress data constitutes a Markov chain [12].

Markov chains illustrate many of the important ideas of stochastic process in an elementary setting. Since the technical requirements are minimal, a relatively complete mathematical treatment is feasible. A Markov chain is completely defined by its one-step transition probability matrix and the specification of a probability distribution on the state of the process at time t = 0. The analysis of Markov chain concerns mainly the calculation of the probabilities of the possible realizations of the process. Central in these calculations are the n-step transition probability matrices $P^{(n)} = \| P_{ij}^{n} \|$. Here $P_{ij}^{n}$ denotes the probability that the process goes from state $i$ to state $j$ in $n^{th}$ transition. Formally,

$$P_{ij}^{(n)} = \Pr\{X_{m+n} = j \mid X_m = i\}$$

The methodology used in this procedure involves generation of the transition matrix from one dimensional time series data. The algorithm for the generation of the transition matrix is discussed below:

1. At first, the time series data sequence $\{t_i\}_{i=1}^{n}$ is normalized to $\{t_i'\}_{i=1}^{n}$, so that each sample value lies in the interval [0,1]. Now starting with this normalized sequence the interval [0,1] is divided into m equal subintervals $I_k, k = 1,2,\cdots,m.$ The value of m is taken as the integer part of $\left(\frac{1}{\sigma}\right)$, where $\sigma$ is the standard deviation of the data set.
2. Generate a new sequence $\{X_i\}_{i=1}^{n}$ where $X_i = k$ if $t_i'$ lies in the interval $I_k$.
3. Generate an $m \times m$ transition matrix $P$ by estimating

$$P_{jk} = \Pr(X_{i+1} = k \mid X_i = j)$$

Interpretation of the characteristics of the chains whose transition matrices are now in hand requires several theoretical results on finite Markov chains. There are plenty of literatures describing the theory of Markov chains. For this application, we will be primarily interested in the fundamental properties of the transition matrix.

If any state j of the Markov chain can be reached starting from any other state i after a finite number of time steps n, i.e. $P_{ij}^{n} > 0$, state i and j communicate, i.e.

$i \to j$ for $i, j \in S$, where $S$ is the set of all possible states of the Markov chain. For $i \to j$ and $j \to i$, states i and j intercommunicate $(i \leftrightarrow j)$. A Markov chain is said to be irreducible if for all $i, j \in S$ it has $i \leftrightarrow j$. By investigating the transition matrix, we observe that the condition $i \leftrightarrow j$ is maintained for all the states. So the Markov chain constructed from the PLC time series data is an irreducible one. Another such important property which can be derived from the transition matrix, is the period of the Markov chain. The period d(*i*) of a state $i \in S$ is defined as

$$d(i) = \gcd\left\{ n \geq 1 : \left(P^n\right)_{i,i} > 0 \right\}.$$

In words, the period of *i* is the greatest common divisor(gcd) of the set of times that the chain can return to *i*, given that we start with $X_0 = i$. If d(*i*) = 1, then the state *i* is said to be aperiodic. A Markov chain is aperiodic if all its states are aperiodic. In our case, the Markov chain appears to be aperiodic. Now any irreducible, aperiodic Markov chain has a unique stationary distribution. A row vector $\pi = (\pi_1, \pi_2, .....\pi_k)$ is said to be a stationary distribution for the Markov chain, if it satisfies

(i) $\pi_i \geq 0,$ for i = 1,2,…..,m and $\sum_{i=1}^{m} \pi_i = 1$ and

(ii) $\pi P = \pi,$ which means $\sum_{i=1}^{m} \pi_i P_{i,j} = \pi_j$ for j =1,2,……..,m.

The experimentally obtained Markov chain converges to such a stationary distribution after a sufficiently long time and one such typical stationary distribution for the strain rate of $1.18 \times 10^{-3} s^{-1}$ is shown in fig. 8.

From the statistical analysis, we find that the dynamics of the type A band propagation is Markovian in nature. Though type A band moves continuously across the specimen, the movement of the band is hindered by the presence of the obstacles like forest dislocations and the band is pinned by the solute atoms. These pinned dislocations are freed by thermal activation and the band moves further. Thus, the band moves jerkily between the obstacles. Now, the dynamics of the band propagation between any two successive pinning positions solely depends on the strength of pinning and the thermal activation associated with the first pinning position between the two. Thus the process is entirely dependent on the current pinning position and does not carry the information about the pinning and strain aging at any earlier position. So the dynamics appears to be memory less which is manifested in the Markovian nature.

**References: -**

**Figure captions: -**

1. A typical segment of drift corrected stress-time curve for the strain rate of $1.18\times10^{-3}s^{-1}$.
2. Autocorrelation function of the drift-corrected stress-time data for the strain rate of $1.18\times10^{-3}s^{-1}$.
3. Running mean for the stress-time data and its surrogates for the strain rate of $1.18\times10^{-3}s^{-1}$.
4. Running standard deviations for the stress-time data and its surrogates for the strain rate of $1.18\times10^{-3}s^{-1}$.
5. Probability distribution of the stress-time series data for the strain rate of $1.18\times10^{-3}s^{-1}$
6. Quantile-quantile plot of the empirical Gaussian data experimental data for the strain rate of $1.18\times10^{-3}s^{-1}$.
7. Normalized eigen value spectrum of the covariance matrix for the strain rate of $1.18\times10^{-3}s^{-1}$.
8. Stationary distribution of the Markov process for the strain rate of $1.18\times10^{-3}s^{-1}$.

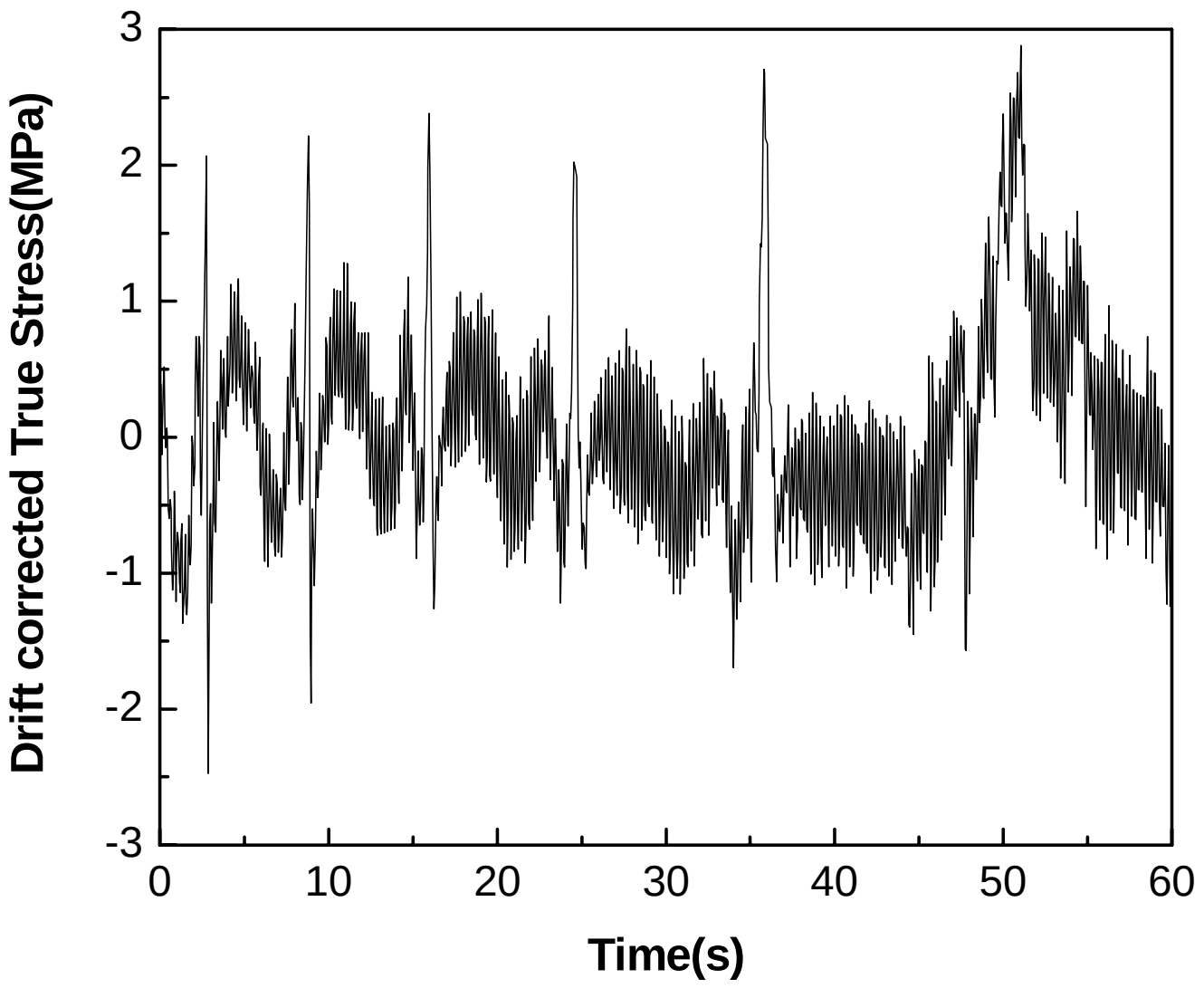

Fig.1

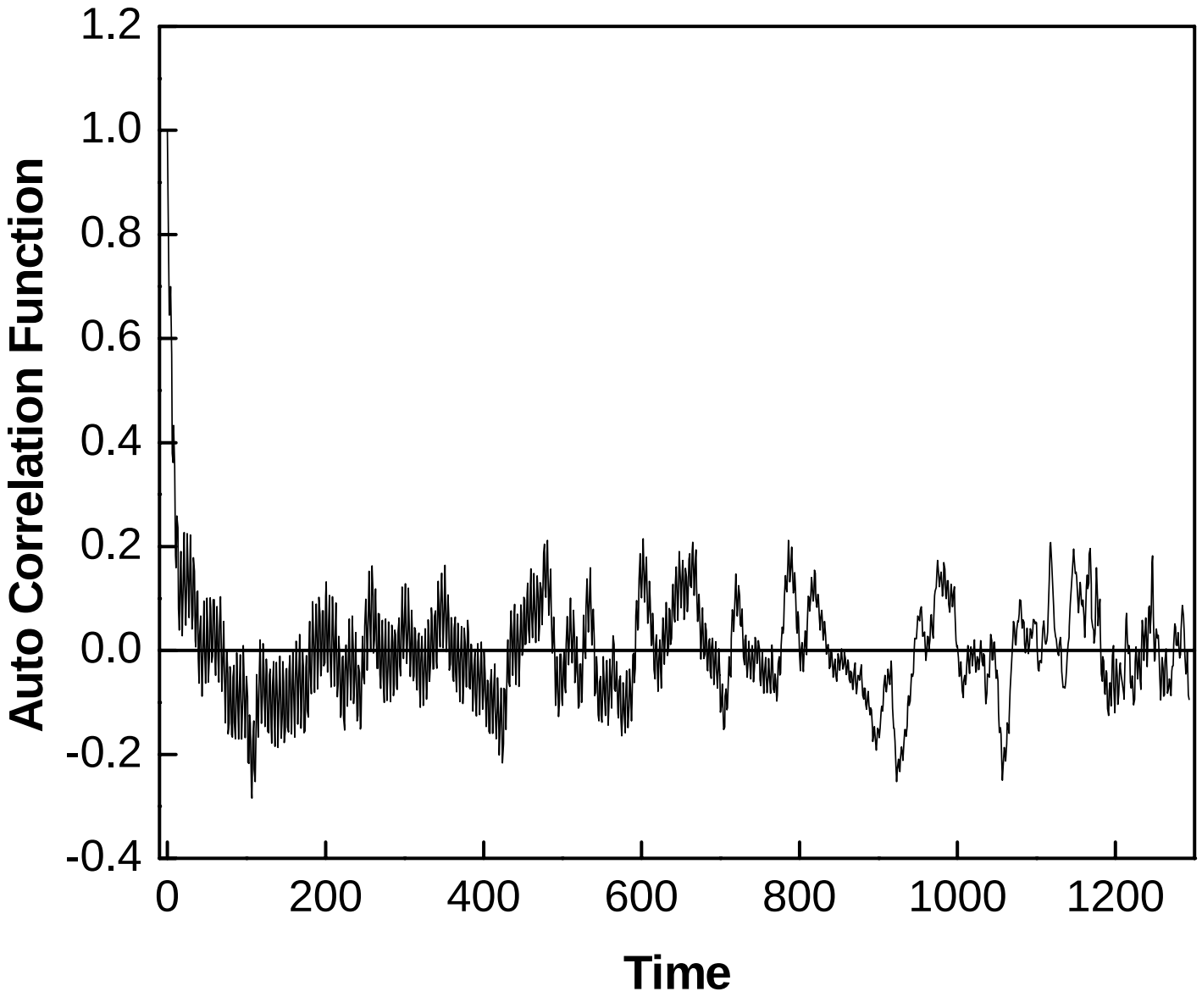

Fig.2

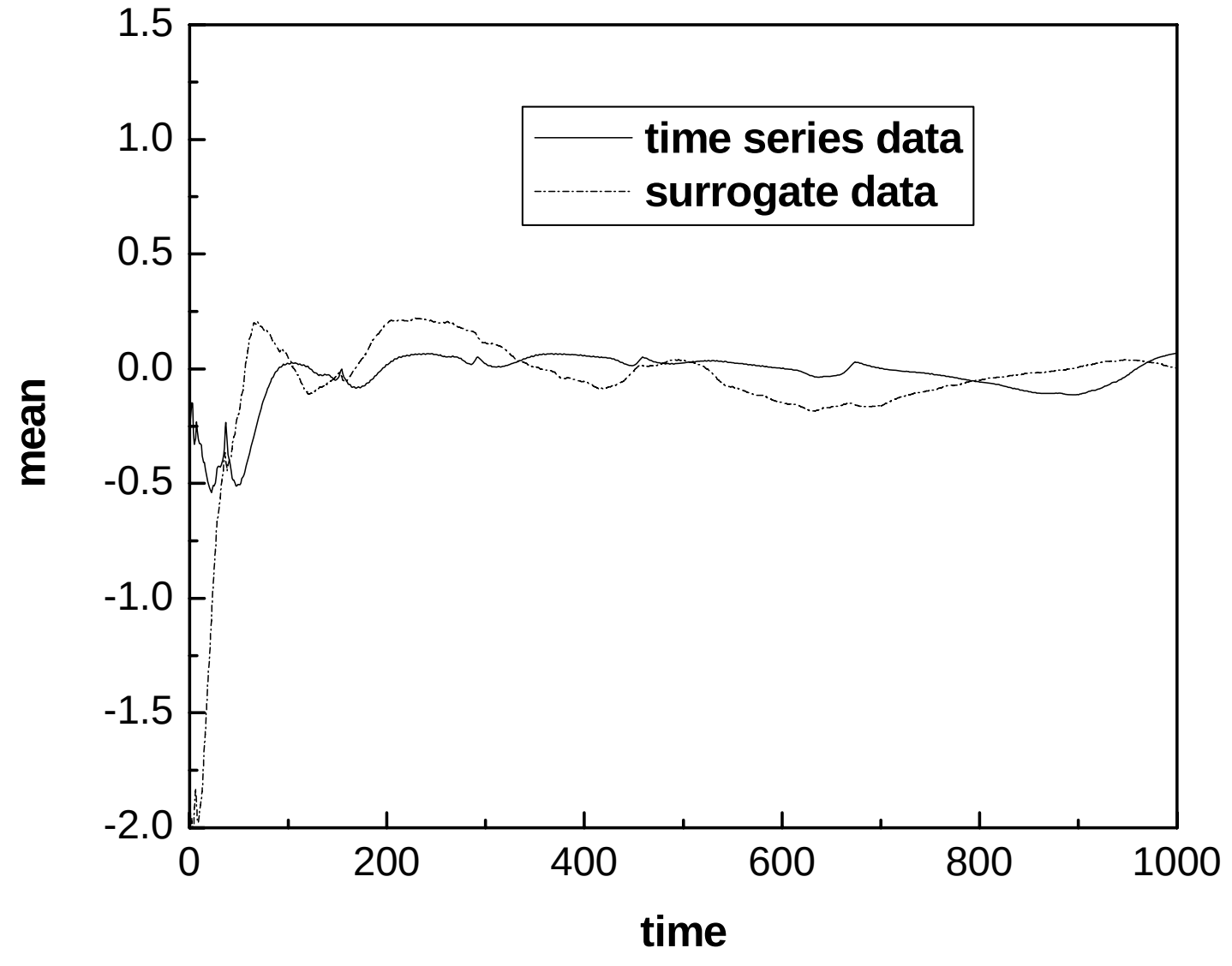

Fig.3

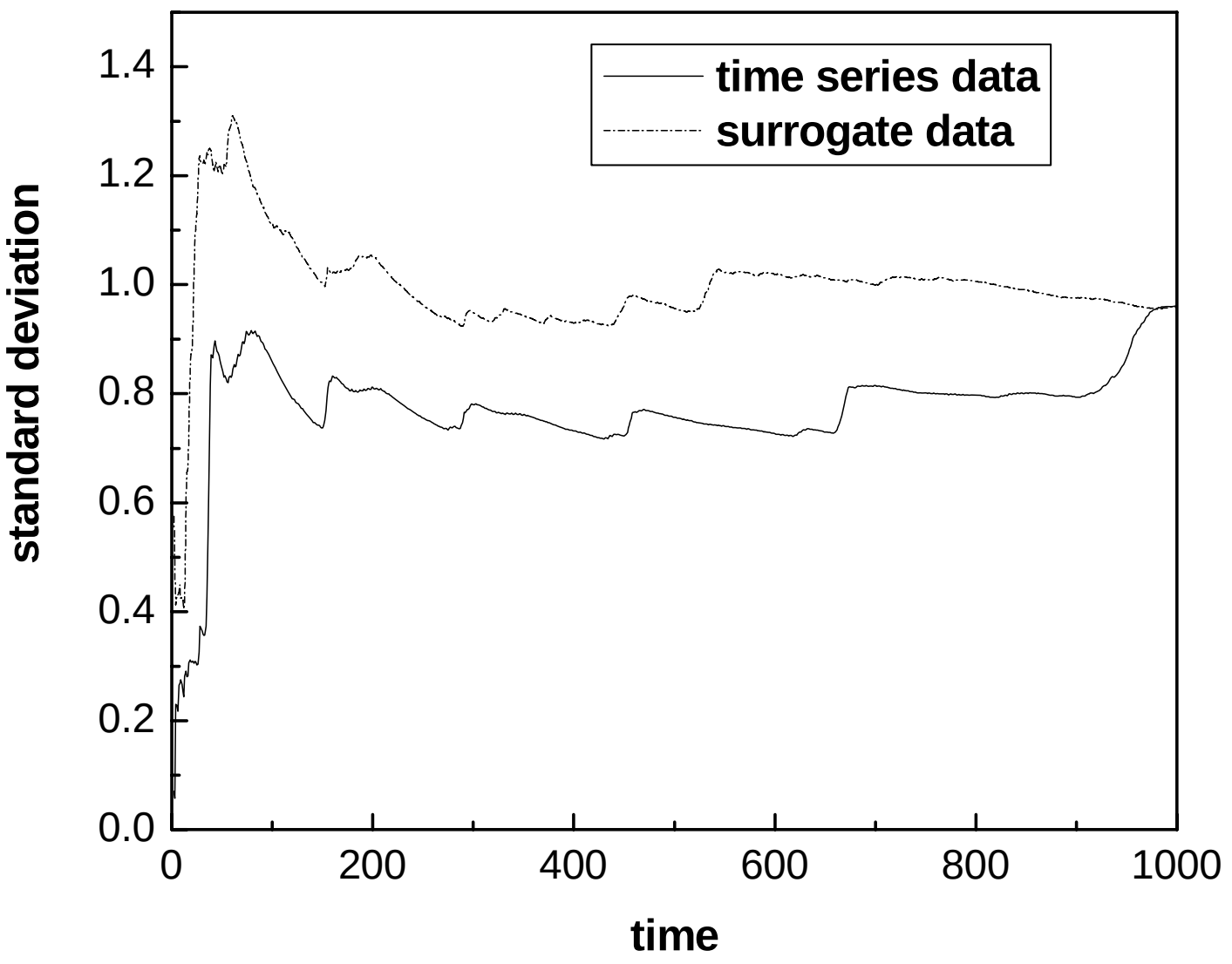

Fig. 4

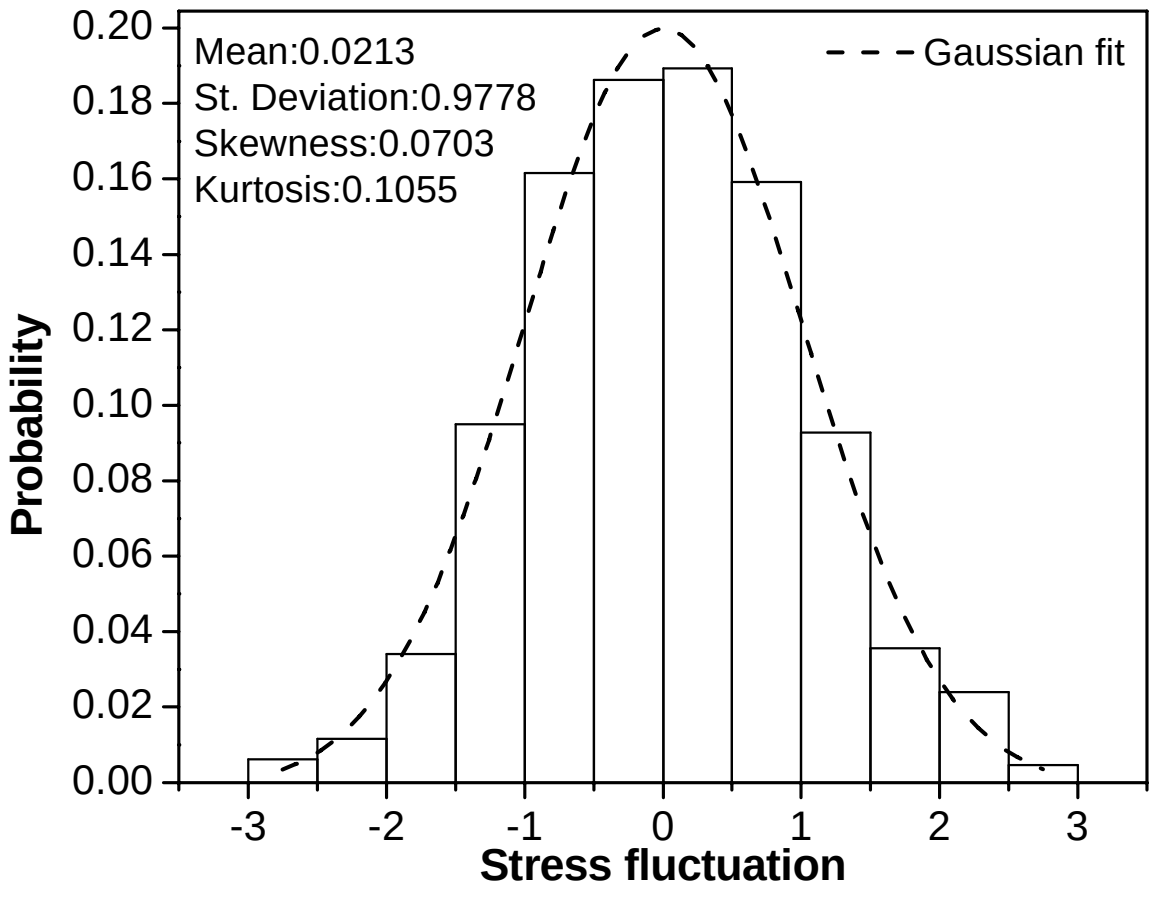

Fig.5

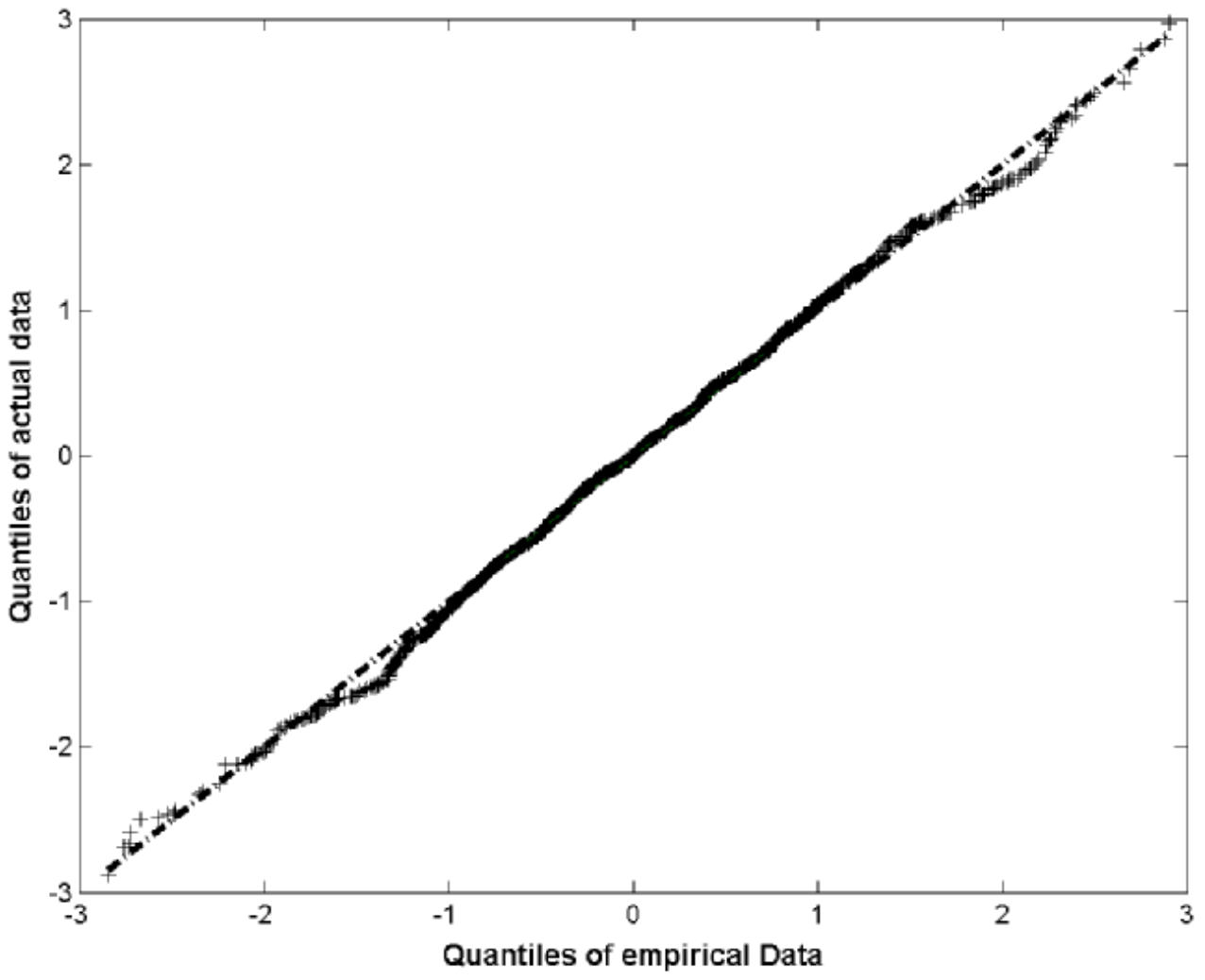

Fig.6

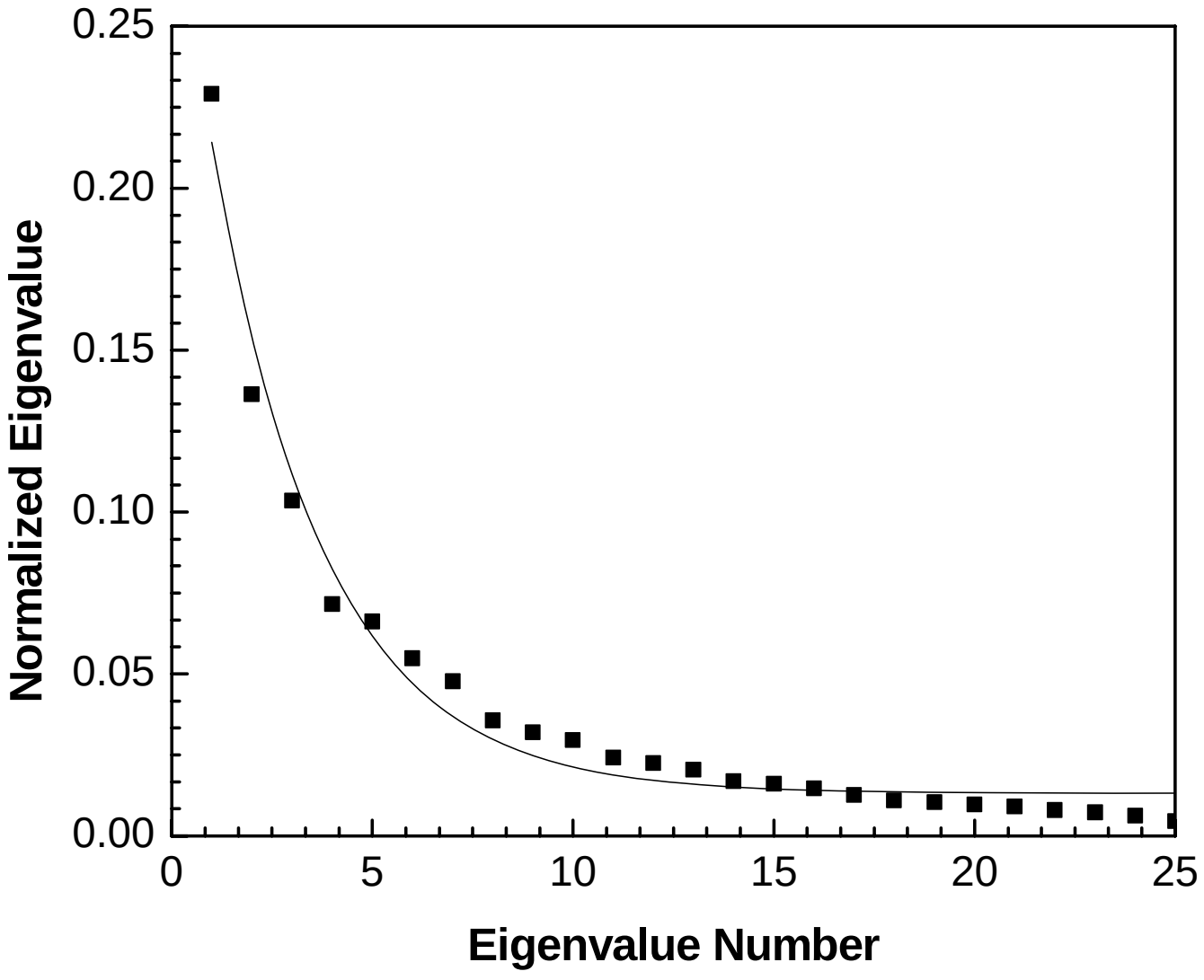

Fig.7

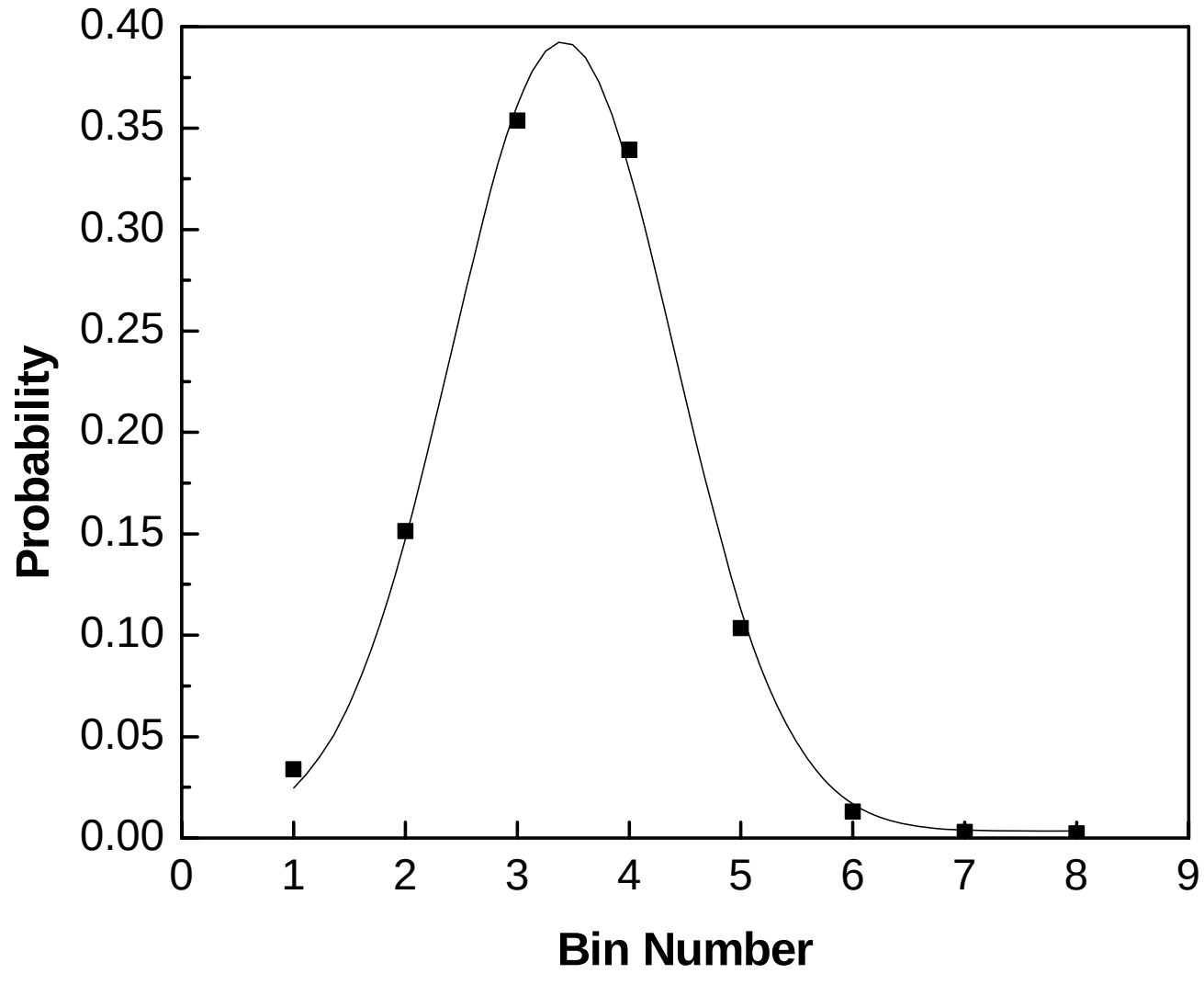

Fig.8